\documentstyle[11pt,aaspp4]{article}
\input psfig.sty

\def\simgt{\lower.5ex\hbox{$\; \buildrel > \over \sim \;$}}
\def\simlt{\lower.5ex\hbox{$\; \buildrel < \over \sim \;$}}
\def\sp{\hspace{1.5pt}}

\lefthead{Vasisht \& Gotthelf}
\righthead{Supernova Remnant Kes\sp73}
\begin{document}

\title{The Discovery of an Anomalous X-ray Pulsar in the Supernova Remnant Kes\sp73}

\author{G. Vasisht}
\affil{California Institute of Technology, MS 105-24, Pasadena, CA 91125}

\author{E. V. Gotthelf\altaffilmark{1}}
\affil{NASA/Goddard Space Flight Center, Greenbelt, MD 20771}

\altaffiltext{1}{Also: Universities Space Research Association}

\begin{abstract}

We report the discovery of pulsed X-ray emission from the compact
source 1E~1841$-$045, using data obtained with the {\it Advanced
Satellite for Cosmology and Astrophysics}. The X-ray source is located
in the center of the small-diameter supernova remnant (SNR) Kes\sp73 and is
very likely to be the compact stellar-remnant of the supernova which formed
Kes\sp73.  The X-rays are pulsed with a 
period of $\simeq 11.8$ s, and a sinusoidal modulation of roughly 30\%.
We interpret this modulation to be the rotation period of an embedded
neutron star, and as such would be the longest spin period for an
isolated neutron star to-date. This is especially remarkable since the
surrounding SNR is very young, at $\sim 2000$ yr old. We suggest that
the observed characteristics of this object are best understood within
the framework of a neutron star with an enormous dipolar magnetic
field, $B \simeq 8\times 10^{14}$ G.

\end{abstract}


\section{INTRODUCTION}

In 1934, Baade and Zwicky published a prophetic paper making a 
phenomenological connection between supernovae (SNs), the core-collapse
of massive stars, and the formation of neutron stars (then hypothetical);
all purely on grounds of energetics. Decades later their conjecture
was first vindicated by the discoveries of young pulsars in the Crab and
Vela supernova remnants (SNRs), and now in a handful of other Galactic
SNR.           
 
Supernova remnants come in at least two distinct
morphological types, i.e., shells and plerions (Weiler \& Sramek
1988), and a majority are the result of core collapse in massive
progenitors (the non-Type Ia SNs; van den Bergh \& Tammann 1991).  The
Baade-Zwicky picture, in its simplest interpretation, is somewhat
problematic.  A majority of SNR appear not to contain either central
pulsars or pulsar plerions (as pulsars are beamed, plerions ought to
be more commonplace than pulsars in the interiors of shells).  The
predominant ``hollowness'' of shell-remnants is ill-understood, and
poses questions about the fate of {\it most} core collapses of massive
stars.  This conundrum is nowhere more apparent than in the studies of
the youngest SNR, especially those of the historical supernovae (Strom
1994).  Of the eight historical supernovae which have expanded into
full-blown SNR, only the Crab Nebula has a pulsar. There is weak
plerionic activity (but no beamed pulsars) in two others (SN 386 A.D.
and SN 1181 A.D.; see Vasisht et al. 1996). It follows, therefore,
that there is a need to give up our notions about the birth properties
of young neutron stars, best typified by the Crab pulsar: fast
rotation ($\simlt 0.1$ s) and a dipole field strength clustered around
$3\times 10^{12}$ G.
That neutron stars may be born in a fashion drastically different from
the Crab has become increasingly evident via recent X-ray studies.
Preliminary evidence of this kind includes: the
discovery of radio-quiet, cooling neutron stars 
(Vasisht et al.  1997 and refs.
therein; Gotthelf, Petre \& Hwang 1997) in SNR, the association of the exotic
soft gamma-ray repeaters with young ($\sim 10^4$ yr old) SNR 
(see Thompson \& Duncan 1995), and
observations of magnetically dominated plerions (Vasisht et al. 1996).
Also, the slowly-spinning ($P \approx 7$ s), anomalous X-ray
pulsar (AXP) in the $\sim 10^4$ yr old SNR CTB 109, has been known for
several years (Gregory \& Fahlman 1980), although its nature is still
widely debated (Mereghetti \& Stella 1995; van Paradijs, Taam \&
van den Heuvel 1995;
Thompson \& Duncan 1996).
  
This {\it Letter} discusses 1E\sp1841$-$045, an unresolved Einstein
point-source discovered near the geometrical center of the shell-type
SNR Kes\sp73 (Kriss et al. 1985). The refined Rosat HRI location of
the object is $\alpha_{J2000} = 18^h41^m19.2^s$ and $\delta_{J2000} =
-04^\circ 56'12.5''$ ($\sim 3''$ at 90\% confidence; (Helfand et al. 1994).
The SNR shows no evidence for an extended plerionic core from either 
radio brightness morphology, polarization properties or spectral
index distribution. This suggests that Kes\sp73, in spite its inferred
youth, lacks a bright radio plerion. To date, no optical counterpart to
the central X-ray source has been identified for 1E\sp1841$-$045.
Herein, we present the discovery of $\sim 12$ s pulsed X-rays 
from 1E\sp1841$-$045 and argue that the source is young and unusual.
In our
companion paper (Gotthelf \& Vasisht 1997; hereafter GV97) we present
the results of imaging-spectroscopy of Kes\sp73 and the compact source.

\section{OBSERVATIONS}

Kes\sp73 was observed with the ASCA Observatory (Tanaka, Inoue \& Holt 1994)
on 1993, October 11-12, as a Performance and Verification (PV) target.
Data were acquired by the two gas imaging spectrometers (GIS2 and
GIS3) and collected with a photon time-of-arrival resolution of $488$
$\mu$s in medium bit-rate mode and $64$ $\mu$s in high bit-rate mode.
We used data made available in the ASCA public archive, screened with the
standard REV1 processing to exclude time intervals corresponding to
high background contamination, i.e., from Earth-block, bright Earth,
and SAA passages. An effective exposure of $\simeq 3.5 \times 10^4$ s
was achieved with each detector and the on-source measured count rates
for the GIS2 and GIS3 instruments were 1.55 (GIS2) and 1.66 (GIS3)
counts s$^{-1}$, respectively.  Here, we concentrate on the GIS data
exclusively and present data from the two Solid-state Imaging
Spectrometers (SIS) on-board ASCA, in our companion paper (GV97), which
reports on spectral and imaging results. We summarize the spectra 
pertinent to this paper below.

The spectrum of 1E\sp1841$-$045 is fit by an absorbed, soft
power-law of photon index $\Gamma \simeq 3.0 \pm 0.2$ ($S_\nu \propto
\nu^{-\Gamma}$). The foreground absorption towards Kes\sp73 is found
to be $N_H \simeq 2 \times 10^{22}$ cm$^{-2}$, and is consistent
with the kinematic distance estimate of 7 kpc (Sanbonmatsu \& Helfand
1992).
The power-law spectral normalization is found to be consistent with no
long term spectral variation when compared to the count rates
observed by Rosat HRI, $\simeq 0.02$ cps (Helfand et al. 1994). We deduce an
unabsorbed model flux of $6.3 \times 10^{-11}$ erg cm$^{-2}$ s$^{-1}$
(0.5 - 10.0 keV) yielding a source luminosity, $L_X \simeq 3.5
\times 10^{35}d_7^2$ erg s$^{-1}$; the SNR distance is $7.0d_7$ kpc.
 

The long term temporal variability of 1E\sp1841$-$045 has been
examined by Helfand et al. (1994) who found no concrete evidence for
flux changes on the 10 year baseline between Einstein and Rosat.  We
examined the ASCA data for aperiodic variability by selecting source
photons from a $5'$ radius aperture and binning them on $\sim 96$ min
(ASCA orbital period), 10 min and 1 min durations. The obtained light curves
were $\chi^2$ tested against a uniform model, but no evidence for significant
variations on these time-scales was found.

A search for coherent pulsation from the central object was made by
combining the two GIS high time resolution datasets ($t_s = 488$
$\mu$s). Photons were selected from the entire SNR region of $\simeq 5'$,
centered on the compact object from (i) the entire energy band (0.5 -
10.0 keV) (ii) the hard band (2.5 - 10.0 keV). The photon
time-of-arrivals were barycentered and binned with resolutions of 488~$\mu$s,
32~ms and 0.5~s. The barycentering and binning procedures were
tested on a series of datasets of the Crab pulsar and PSR
0540$-$69. Fourier transforms on the entire datasets were performed at
each time resolution, interesting periodicities were harmonically
summed and later folded at the period of interest. 
 
A significant high-Q X-ray modulation with no overtones
is seen at a period $P \simeq 11.766684$ s ($f = 0.08498571 \pm
0.00000004$ Hz; see Figs 1 and
2).  The modulation is obvious in all the above datasets and
separately in either GIS on-source time series; it is not observed in
off-source GIS data, making it unlikely to be an instrumental
artifact. The period emerges with greatest significance in time-series
which contain emission mainly from the central source (hard-band, 2.5
- 10.0 keV), and is not significant in the soft energy band between
0.5 - 2.5 keV, where the nebula is dominant.  We suggest that the
central object is weakly pulsed at 11.7667 s, possibly a neutron star
spin period. The pulsed luminosity is $L_X \approx 5\times
10^{34}d_7^2$ erg s$^{-1}$, and the modulation level is about 35\% of the
steady flux from the compact source after subtraction of the estimated
contribution of the 
background and the SNR thermal component above 2.5 keV.

With a period in hand, we have reanalyzed the 18-ks of Rosat HRI data
obtained between March 16-18, 1992 (Helfand et al. 1994), selecting the
few ($\simeq 650$) source and background photons available from the
vicinity of 1E\sp1841$-$045.  We perform a conditional search in a
small range of periods around 11.76 s using the folding technique (12
phase bins per fold); the resulting periodogram is displayed in Fig
1b. We cautiously forward the suggestion of a peak-up at a barycentered
period of 11.7645 s (4.0 $\sigma$; which is the expected significance
given the HRI count rate and the ASCA derived modulation); the observed
FWHM of the periodogram excess of $\simlt 10^{-3}$ s is consistent
with expectations, given the period and the time-span of the HRI
observations.  Also, the crest and trough in the HRI profile (Fig 2b)
match those of the GIS profile quite well.  A linear interpolation,
assuming steady spin-down, gives a period derivative of $\dot P \simeq
4.73 \times 10^{-11}$ s s$^{-1}$.

\section{DISCUSSION}

Our interpretation of 1E\sp1841$-$045 is that of a young neutron
star that was born during a supernova that now forms the SNR Kes\sp73.
The kinematic distance to the SNR of 6.7 - 7.5 kpc is consistent
with its high foreground X-ray absorption, $N_H \simeq 2\times 10^{22}$
cm$^{-2}$. The small shell radius ($R_s \simeq 4.7d_7$ pc), along with
intense radio and X-ray shell emission are characteristics of a young
supernova remnant. This notion is supported by our detection
of a hot thermal ($kT \simeq 0.8$ keV) X-ray continuum in the shocked
gas. Helfand et al. (1994) have argued that the SNR is likely  to
be in transition between free expansion and the adiabatic phases, whereby,
the Sedov age of $\tau_s < 2500$ yr must be an upper bound. The SIS
spectra show the enhancement of Mg over the species S, Si and Fe,
along with possible evidence for highly absorbed emission from O and
Ne in the 0.5-0.9 keV range (see GV97). During the course of their
evolution, massive stars produce large quantities of O-group (O, Ne
and Mg) which are ejected during the supernova (see Hughes et al. 1994). 
Hence, there is 
spectral evidence that Kes\sp73 is young and still ejecta dominated.
On spectral grounds we also favor a Type II or Ib origin for Kes\sp73,
i.e., from a massive progenitor. We consider it unlikely that the
neutron star was born in an accretion-induced collapse of a heavy
white dwarf (Lipunov \& Postnov 1985).

The pulsar in 1E\sp1841$-$045 has common properties with the
peculiar X-ray pulsar, 1E\sp2259+586 (Gregory \& Fahlman 1980; Corbet
et al. 1995), and the soft gamma-ray repeaters (Thompson \& Duncan
1996).  1E~2259+586 is a 7-s spin rotator, and coincides with a
$\simgt 10^4$ yr-old SNR (CTB 109). Much like 1E\sp1841$-$045, it has
a soft X-ray spectrum best represented by a blackbody at 0.45 keV with
a non-thermal tail with $S_\nu \propto \nu^{-3}$, and $L_x \simeq 0.5
- 1\times10^{35}$ erg s$^{-1}$.  It has a history of nearly steady
spin-down and no detected binary modulation (Iwasawa, Koyama \&
Halpern 1992), optical companion, or quiescent radio emission. There
are four other X-ray pulsars, 4U\sp0142+614 (Hellier 1994; Israel et
al. 1994), 1E\sp1048.1-5937 (Seward, Charles \& Smale 1986),
RXJ\sp1838$-$0301 (Schwentker 1994; possibly also associated with a
$\simgt 10^4$ yr old SNR), and 1RXS J170849.0$-$400910 (Sugizaki et
al. 1997) which have low luminosities ($L_X \sim 10^{35} - 10^{36}$
erg s$^{-1}$), periods of order 10-s that are steadily increasing,
soft spectra and no detected companions or accretion disks. In all
the above cases, spectra can be fit with soft power-laws with indices
in the range 2.3 - 3.5 (Corbet et al. 1995). Collectively, these have
been grouped into a class called the breaking X-ray pulsars
(Mereghetti \& Stella 1995) or alternatively, the anomalous X-ray
pulsars (van Paradijs et al. 1995).  We observe AXPs through a
substantial distance in the Galactic disk (with foreground columns
densities $\sim 10^{22}$ cm$^{-2}$), which suggests that they are not
commonplace.

The rotational energy of 1E\sp1841$-$045 is far too small to power
its total X-ray emission of $L_X \sim 4 \times 10^{35}d_7^2$. The maximum
luminosity derivable just from spin-down is 
$$ L_X \approx 4\pi^2I {\dot P \over P^3} \sim {1\over 2\tau_s} I
({2\pi\over P})^2 \sim 10^{33}~\hbox{erg s$^{-1}$},$$ where $I$ is the
moment of inertial of the neutron star, and $\tau_s \sim 2\times 10^3$
yr is the SNR age. The mechanisms for powering the X-rays could then
be either (i) accretion from a high-mass X-ray binary (Helfand et
al. 1994), a low mass companion (Mereghetti \& Stella 1995), or a
fossil disk (van Paradijs et al. 1995), or a merged white
dwarf (Paczyncski 1990) (ii) intrinsic energy loss,
such as initial cooling or the decay of magnetic fields in a magnetic
neutron star (Thompson \& Duncan 1996).

The strongest argument for accretion as
the source of energy is that the inferred accretion rate is just that
required if the NS is close to its equilibrium spin period
$P_{eq}$, with a field $B \sim 10^{12}$~G typical for young pulsars
(see Bhattacharya \& van den Heuvel 1991)
$$ P_{eq} \simeq 10~\hbox{s}~({B_d\over
5\times10^{11}~\hbox{G}})^{6/7}({L_X\over 10^{35}~\hbox{erg
s$^{-1}$}})^{-3/7}.$$ 
However, only a pathological evolution scenario involving accretion
could bring an energetic dipole rotator to its present rotation rate
within $\simeq 2\times 10^3$ yr. 
The strongest support for 1E\sp1841$-$045 as an
accretor will be from future identification of an
infra-red (large foreground $A_V \simeq 10$ mag.) companion or an
accretion disk. As in the case of
other AXPs an infrared counterpart may not be easily
identified (Coe, Jones \& Lehto 1994 and refs. therein).
The pulsar has properties  that may already preclude accretion as a power
source: (i) high-mass neutron star binaries with a NS 
accreting from the companion wind sometimes go into low luminosity states with 
$L_X \sim 10^{35}$ erg s$^{-1}$, and have periods in the range 0.07 - 900
s. However, in general they display hard spectra ($0.8 < \Gamma <
1.5$) and strong aperiodic variability on all time-scales
(Nagase 1989). If the X-ray source is indeed a high mass X-ray
binary then the inferred $\dot P$ could be the result of
orbital Doppler effects. 
(ii) Neutron stars with disk accretion from a low mass
companion or a fossil disk, with the latter having formed from SN
debris or a Thorne-Zytkow phase (van Paradijs et al. 1995), 
should display similar accretion noise in the light-curve. 
(iii) Other AXPs show near steady spin-down on time-scales $10^4 - 10^5$
yr, although this is controversial (Baykal \& Swank 1996; Corbet
\& Mihara 1997). The long term torque behavior of 1E\sp1841$-$045 will only
be evident with future observations.
iv) Finally, accretion models would have to be stretched in order to explain the
slow rotation period (inside a young SNR) 
and its associated spin-down time-scale, $\simlt 2500$
yr. First, it difficult for accretion torques to spin-down a pulsar
to 12-s in $\sim 10^3$ yr from initial periods $P_i \simlt 10^2$ ms
unless, of course, the pulsar were {\it born a very
slow rotator}, which is quite interesting in its own right.  Secondly,
if the pulsar were rotating near its equilibrium period, as in the Ghosh
and Lamb (1979) scenario, the spin-down time of ${P/ 2\dot
P} \sim 3900$ yr is inconsistent with the implied accretion rate,
$\dot M \simeq 10^{-11}$ M$_\odot$ yr$^{-1}$ (assuming the pulsar has
a normal dipolar field $\simeq 10^{12}$ G); these usually lie in
range $10^4 - 10^5$ yr.
 
A dipolar magnetic field vs. period scaling can also be obtained under
the assumption that the neutron star is isolated, and has undergone
conventional pulsar spin-down from torques provided by a relativistic
wind, as in the Crab. For dipolar secular spin-down, the implied
$B_{dipole}$ is enormous,
$$ P = 10~\hbox{s}~({t\over 3\times
10^3~\hbox{yr}})^{1/2}({B_{dipole}\over 10^{15}~\hbox{G}})({R \over
15~\hbox{km}})^2({M \over 1.4~\hbox{ M$_\odot$}})^{-1/2}.$$ Such
highly magnetized neutron stars (with dipolar field strengths $B
\approx 10^{14} - 10^{15}$ G) or ``magnetars'' have been postulated by
Thompson \& Duncan (1995 and refs. therein) to explain the action of soft
gamma-ray repeaters.  Magnetars have magnetic flux densities that are a
factor $\simgt 10^2$ larger than the typical $B \sim 10^{12}$ G fields
supported by radio or X-ray pulsars and perhaps represent a tail of
$B$-field distribution in NS.  After birth, they spin-down too rapidly
to be easily detectable as radio pulsars, assuming that they are
capable of radio pulsar action at all. The dipole energy in the star's
exterior, a small fraction of the total magnetic energy, exceeds the
rotational energy of the NS after roughly $2 B_{15} ^{-4}$ yr, where
$B = 10^{15}B_{15}$ G. Magnetism then quickly becomes the dominant source 
of free energy in an isolated magnetar.

The derived spin-down age of the pulsar $\sim 3900$ yr, is consistent
with the SNR age (ages inferred from the estimator $P/2\dot P$ are
larger than the true age as they measure linear spin-down).The
equivalent dipolar field is $B_{dipole} \simeq 3.2\times10^{19}~(P\dot P)^{1/2}
\approx 8\times 10^{14}$ G.  There is an intriguing possibility
that the the pulsar in Kes\sp73 was born as a magnetar $\sim
2\times10^3$ yr ago, and has since spun-down to the long period of
11.7-s due to rapid dipole radiation losses. It could be unobservable
as a radio pulsar due to period dependence of 
beaming (Kulkarni 1992); it is also possible the radio pulsar mechanism may
operate differently or not at all above the quantum critical field,
$B_{cr} \simeq 4.4\times 10^{13}$ G. In a magnetar, the observed X-ray
luminosity would be driven by the decay of the stellar B-field via
diffusive processes, which set in at an age $\sim 10^3 - 10^4$ yr
(Thompson \& Duncan 1996; we assume that diffusion of field lines through the crust
by Hall drift, and the core by ambipolar diffusion occur on
time-scales $\tau_{Hall} \simeq 5\times 10^8 B_{12}^{-1}$ yr and
$\tau_{amb} \simeq 3\times 10^9 B_{12}^{-2}$ yr, respectively;
Goldreich \& Reisenegger 1992).  Magnetic field decay powers the star
on average, at a steady rate of $(1/6t_d)R^3B_{dipole}^2 \sim 10^{36}$
erg s$^{-1}$ for the first decay time, $t_d \sim 10^3$ yr.  Decay in
the core is likely to keep the stellar surface hot via release of
internal magnetic free energy, while crustal decay is likely to set up
a steady spectrum of Alf\'ven waves in the magnetosphere which
can accelerate particles to produce the soft non-thermal tail observed
in the pulsar spectrum (GV97). Such a neutron star may in time
($\sim 10^4$ yr) display the soft gamma-ray repeater phenomenon
(Thompson \& Duncan 1996).

In conclusion, we reiterate that 1E\sp1841-045 is in our estimation a
young ($\sim 2000$ yr-old)
neutron star spinning at an anomalously slow rate of $\simeq 11.8$-s,
possibly with very strong torques on its rotation.  The claim of rapid
spin-down is based on a weak detection of periodicity in the archival
Rosat HRI data, and needs to be urgently tested in future
observations. Whatever the final consensus on 1E\sp1841$-$045, it is a
most unusual and exciting object, the understanding of whose nature
should make us rethink important aspects about the birth process of 
neutron stars as a whole.

\noindent
{\bf Acknowledgments:} First of all, we thank the HEASARC archives
for making the ASCA data available to us. GV would like to thank
Shri Kulkarni for discussions and for making the trip to GSFC
possible, and to the LHEA at GSFC for hosting him. 
We thank the LHEA for generous use of its facilities.
GV's research is supported by NASA and NSF grants. 
EVG's research is supported by NASA. GV thanks David Helfand for 
earlier discussions on Kes 73.

\clearpage

\begin{figure} 

\noindent
{\bf Fig. -- 1} (Top) A power spectrum of photons from both GIS cameras displayed
in the range 11.75 - 11.78 s. Photons were selected from the hard-band
(2.5 - 10.0 keV). The main peak near 11.766684 s is the putative pulsar period. 
The powerful side-lobe peaks are separated from the main peak by 0.00017 Hz,
the ASCA orbital period. (Bottom) A periodogram of $\chi^2$ vs. period of the HRI data,
peaks up at roughly 4$\sigma$ as is expected from the GIS profile 
(even though the energy range is different). The peak-up period is 11.7645 s. 
The total number of HRI counts for this was $\sim 800$. The search was done with
twelve bins across the folding period.

 \bigskip

\noindent
{\bf Fig. -- 2} (Top) A normalized folded profile of the GIS data
(including background),
with 12 bins of resolution, and a folding period of 11.7667 s. The
profile is roughly sinusoidal, with about $\sim$ 35\% modulation (after accounting
for the background). The start epoch of folding is MJD $4.9271609362\times
10^4$.
(Bottom) A normalized folded profile of Rosat HRI data, with 12 bins
of resolution, at a folding period of 11.7645 s. The start epoch of 
folding is MJD $4.8697189\times 10^4$.

\end{figure}

\clearpage
\begin{figure}
\centerline{\psfig{file=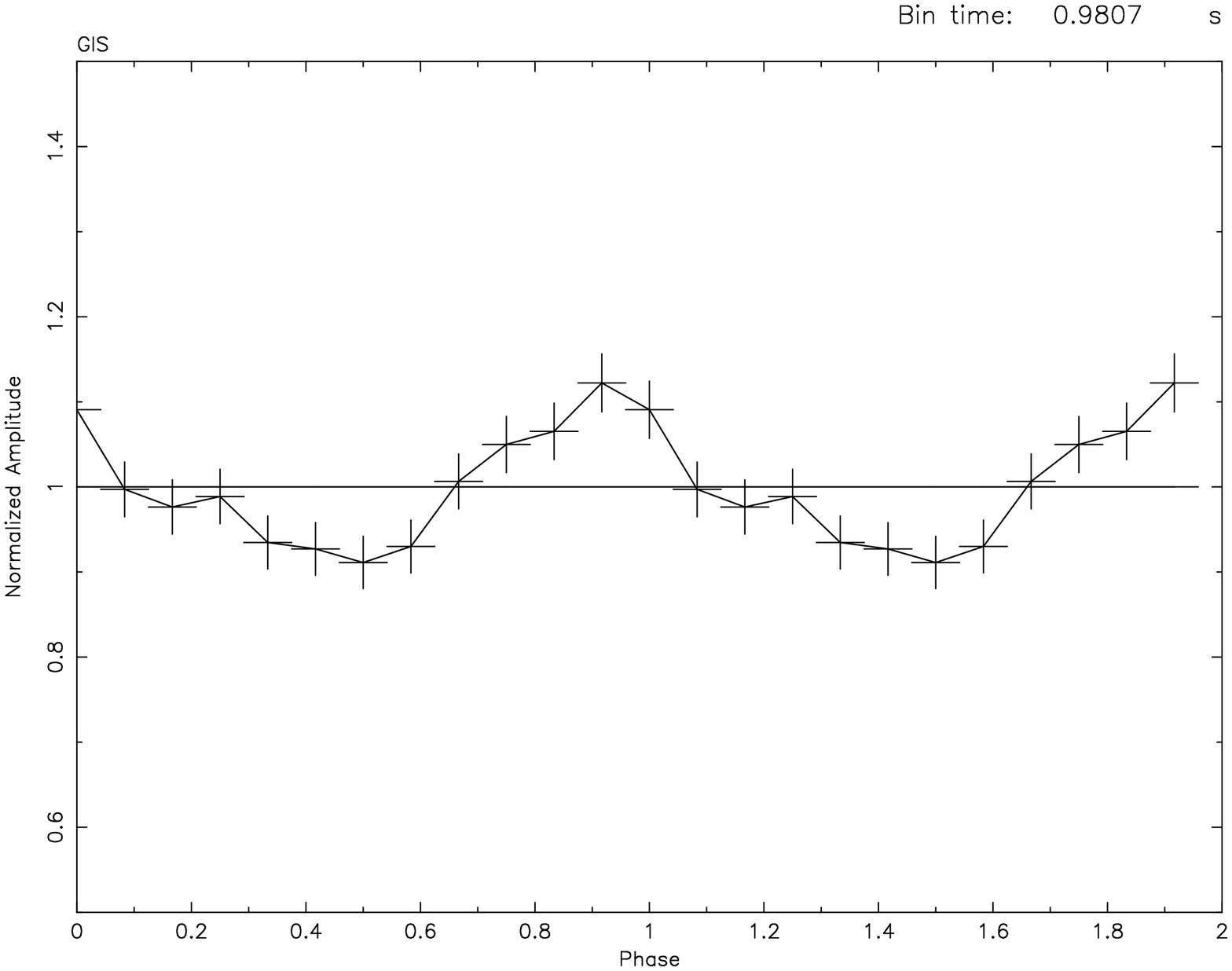,width=6in,angle=270}}
\centerline{\psfig{file=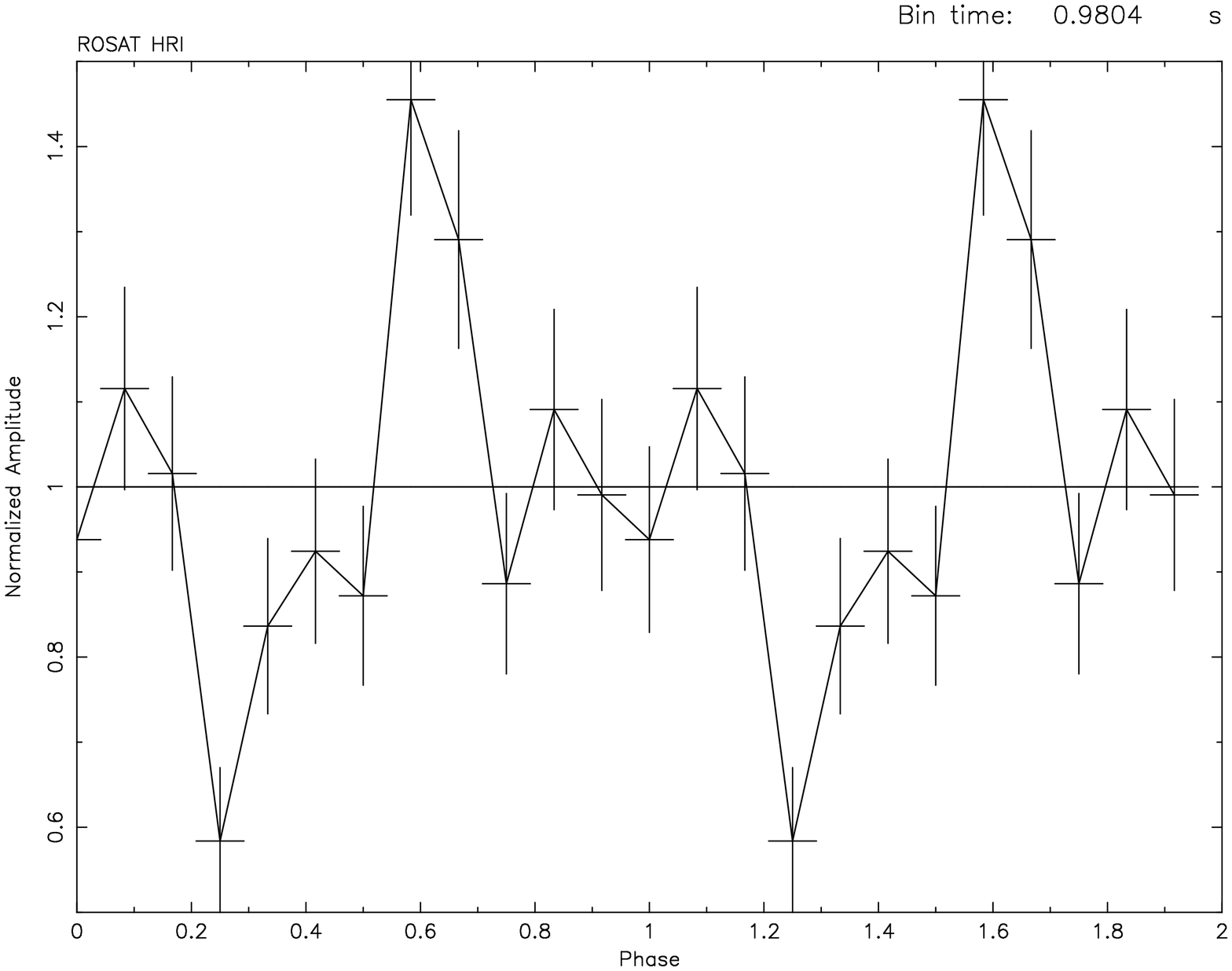,width=6in,angle=270}}
\end{figure}

\clearpage
\begin{figure}
\centerline{\psfig{file=gisallpower.ps,width=4in,angle=270}}
\centerline{\hskip 0.0in \psfig{file=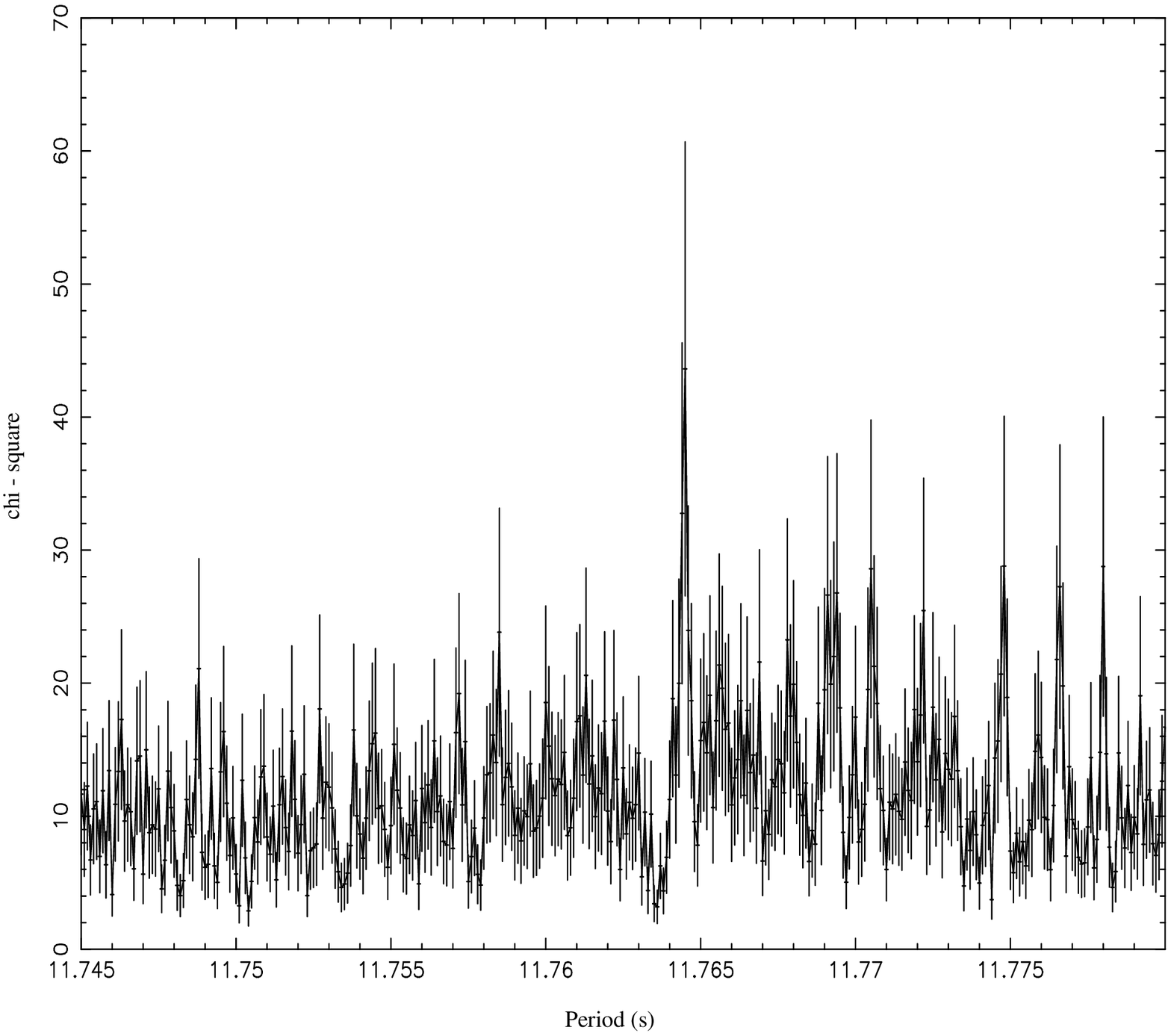,width=4.0in,height=3in,angle=270}}
\end{figure}



\begin{thebibliography}{}

\bibitem[Baade \& Zwicky 1934]{Baa34} Baade, W. \& Zwicky, F. 1934, Phys.
Rev., 45, 138

\bibitem[Baykal \& Swank 1996]{Bay96} Baykal, A. \& Swank, J. 1996,
ApJ, 460, 470

\bibitem[Bhattacharya \& van den Heuvel 1991]{bha91} Bhattacharya, D.
\& van den Heuvel, E. P. J. 1991, Phys. Rep., 203, 1

\bibitem[Coe, Jones \& Lehto 1994]{Coe94} Coe, M. J., Jones, L. R. \&
Lehto, H. 1994, MNRAS, 270, 178

\bibitem[Corbet et al. 1995]{Cor95} Corbet, R. H. D., Smale, A. P., Ozaki,
M., Koyama, K. \& Iwasawa, K. 1995, ApJ, 443, 786

\bibitem[Gregory \& Fahlman 1980]{Gre80} Gregory, P. C. \& Fahlman, G. G.
1980, Nature, 287, 805

\bibitem[Goldreich \& Reisenegger 1992]{gol92} Goldreich, P. \&
Reisenegger, A. 1992, \apj, 395, 250

\bibitem[Gotthelf, Petre, \& Hwang 1997]{Gph97} Gotthelf, E. V., Petre, R. \&
Hwang, U. 1997, ApJL, submitted

\bibitem[Gotthelf \& Vasisht 1997]{Got97} Gotthelf, E. V. \&
Vasisht, G. 1997, ApJL, submitted (GV97)


\bibitem[Helfand et al. 1994]{Hel94} Helfand, D. J, Becker, R. H., 
Hawkins, G. \& White, R. L.  1994, ApJ, 434, 627

\bibitem[Hellier 1994]{Heli94} Hellier, C. 1994, MNRAS, 271, L21

\bibitem[Hughes et al. 1994]{Hug94} Hughes, J. P., Hayashi, I., 
Helfand, D., Hwang, U., Itoh, M., Kirshner, R., Koyama, K.,
Markert, T., Tsunemi, H. \& Woo, J. 1994, ApJ, 444, L81 

\bibitem[Iwasawa et al. 1992]{Iwa92} Iwasawa, K., Koyama, K. \& Halpern
J. P. 1992, ApJ, PASJ, 44, 9

\bibitem[Israel et al. 1994]{Isr94} Israel, G. L., Mereghetti, S. \&
Stella, L. 1994, ApJ, 433, L25

\bibitem[Kriss et al. 1995]{Kri85} Kriss, G. A., Becker, R. H., Halfand,
D. J. \& Canizares, C. J. 1985, ApJ, 288, 703

\bibitem[Kulkarni 1992]{Kul92} Kulkarni, S. R. 1992, Phil. Trans. R.
Soc. London, A, 341, 77

\bibitem[Lipunov \& Postnov 1985]{Lip85} Lipunov, V. M. \& Postnov, K. A.
1985, A\&A, 144, L13

\bibitem[Mereghetti \& Stella 1995]{Mer95} Mereghetti, S. \& Stella, L.
1995, ApJ, 442, L17

\bibitem[Nagase 1989]{Nag89} Nagase, F. 1989, PASJ, 41, 1

\bibitem[Paczynski 1990] Paczynski, B. 1990, ApJ, 365, L9

\bibitem[Sanbonmatsu \& Helfand 1992]{San92} Sanbonmatsu, K. Y. \&
Helfand, D. J. 1992, AJ, 104, 2189

\bibitem[Schwentker 1994]{Sch94} Schwentker, O. 1994, A\&A, 286, L47

\bibitem[Seward et al. 1986]{Sew86} Seward, F. D., Charles, P. A.
\& Smale, A. P. 1986, ApJ, 305, 814

\bibitem[Strom 1994]{Str94} Strom, R. G. 1994, A\&A, 288, L1

\bibitem[Sugizaki et al. 1997]{Sug97} Sugizaki et al. 1997, IAUC,
6585

\bibitem[Tanaka et al. 1994]{Tan87}
Tanaka, Y., Inoue, H. \& Holt, S. S. 1994, PASJ, 46(3), L37
 
\bibitem[Duncan \& Thompson 1995]{dt95} Thompson, C., \& Duncan, R. C.
1995, \mnras, 275, 255 (TD95)
 
\bibitem[Duncan \& Thompson 1995]{dt95} Thompson, C., \& Duncan, R. C.
1995, \apj, 473, 322

\bibitem[van den Bergh \& Tammann 1991]{van91} van den Bergh, S. \&
Tammann,  1991, ARA\&A, 29, 363

\bibitem[van Paradijs et al. 1995]{van95} van Paradijs, J., 
Taam, R. E. \& van den Heuvel, E. P. J. 1995, A\&A, 299, L41

\bibitem[Vasisht et al. 1996]{Vas96} Vasisht, G., Aoki, T., 
Dotani, T., Kulkarni, S. R. \& Nagase, F. 1996, ApJ, 456, L59

\bibitem[Vasisht et al. 1997]{Vas97} Vasisht, G., Kulkarni, S. R.,
Anderson, S. B., Hamilton, T. T. \& Kawai, N. 1997, ApJ, 476, L43

\bibitem[Weiler \& Sramek 1988]{Wei88} Weiler, K. W. \&\ Sramek, R. A. 1988, ARAA, 25, 295
 

\end{thebibliography}
\end{document}